\documentclass[sigconf]{acmart}

\AtBeginDocument{%
  \providecommand\BibTeX{{%
    \normalfont B\kern-0.5em{\scshape i\kern-0.25em b}\kern-0.8em\TeX}}}

\setcopyright{acmcopyright}
\copyrightyear{2021}
\acmYear{2021}
\acmDOI{10.1145/1122445.1122456}

\acmConference[Jerusalem '21]{Jerusalem '21: The 14th ACM International Conference
on Web Search And Data Mining}{March 8--12, 2021}{Jerusalem, Israel}
\acmBooktitle{Jerusalem '21: The 14th ACM International Conference
on Web Search And Data Mining, March 8--12, 2021, Jerusalem, Israel}
\acmPrice{15.00}
\acmISBN{978-1-4503-XXXX-X/18/06}

\usepackage{subcaption} 
\usepackage[utf8]{inputenc}
\usepackage{natbib}
\usepackage{graphicx}
\usepackage{enumitem}
\usepackage{array}
\usepackage{algorithm,algpseudocode}
\usepackage{amsmath}
\usepackage{amsfonts}
\usepackage{multirow}
\usepackage{caption}
\usepackage{balance}
\begin{document}

\title{How to Measure Your App: A Couple of Pitfalls and Remedies in Measuring App Performance in Online Controlled Experiments}

\settopmatter{authorsperrow=4}

\author{Yuxiang Xie}
\affiliation{%
  \institution{Snap Inc.}
  \city{Santa Monica}
  \state{CA}
  \country{USA}
}
\email{yxie@snap.com}

\author{Meng Xu}
\affiliation{%
 \institution{Snap Inc.}
  \city{Santa Monica}
  \state{CA}
  \country{USA}}
\email{mxu@snap.com}

\author{Evan Chow}
\affiliation{%
  \institution{Snap Inc.}
  \city{Santa Monica}
  \state{CA}
  \country{USA}}
\email{echow@snap.com}

\author{Xiaolin Shi}
\affiliation{%
  \institution{Snap Inc.}
  \city{Santa Monica}
  \state{CA}
  \country{USA}}
\email{xiaolin@snap.com}

\renewcommand{\shortauthors}{Xie, et al.}

\begin{abstract}

 Effectively measuring, understanding, and improving mobile app performance is of paramount importance for mobile app developers. Across the mobile Internet landscape, companies run online controlled experiments (A/B tests) with thousands of performance metrics in order to understand how app performance causally impacts user retention and to guard against service or app regressions that degrade user experiences. To capture certain characteristics particular to performance metrics, such as enormous observation volume and high skewness in distribution, an industry-standard practice is to construct a performance metric as a quantile over all performance events in control or treatment buckets in A/B tests. In our experience with thousands of A/B tests provided by Snap, we have discovered some pitfalls in this industry-standard way of calculating performance metrics that can lead to unexplained movements in performance metrics and unexpected misalignment with user engagement metrics.
 In this paper, we discuss two major pitfalls in this industry-standard practice of measuring performance for mobile apps. One arises from strong heterogeneity in both mobile devices and user engagement, and the other arises from self-selection bias caused by post-treatment user engagement changes. To remedy these two pitfalls, we introduce several scalable methods including user-level performance metric calculation and imputation and matching for missing metric values. We have extensively evaluated these methods on both simulation data and real A/B tests, and have deployed them into Snap's in-house experimentation platform.

\end{abstract}


\keywords{performance metrics, online controlled experiment, self-selection bias, sample ratio mismatch}

\maketitle

\section{Introduction} \label{sec:intro}


As mobile applications (apps) gain increasing use in everyday life, the question of how to measure, understand, and improve app performance has become ever more important for mobile app developers. In contrast to most traditional online service companies who primarily use performance metrics as guardrails for fundamental reliability and business constraints, mobile app developers aim to improve performance in order to boost user retention and engagement. For instance, from late 2017 there was a large engineering initiative to rewrite the Android version of Snapchat for faster performance, aligning with that of the iOS version, as a means to retain and attract more Snapchat users\footnote{https://www.androidauthority.com/snapchat-redesign-android-jerry-hunter-gustavo-moura-jacob-andreou-interview-971385}. Hence, finding accurate methods to measure app performance, such as tracking how long it takes to open the app or how long it takes to send a message, is of foundational importance for mobile business growth.


There are three major difficulties in measuring mobile app performance. First, due to the snowballing growth of the mobile device industry, there are tens of thousands of distinct mobile devices and models available on the market. Within the Snapchat userbase, for instance, we have seen several thousand different Android mobile devices. This vast heterogeneity in mobile devices, ranging from different hardware configurations to disparate Wi-Fi conditions, makes measuring the performance of even a single mobile app such as Snapchat an extremely complex task. Second, mobile users are not typically evenly distributed across the performance spectrum from lower-end to higher-end devices. For instance, there are substantially more Snapchat users on lower-end Android devices than on higher-end ones, which means lower-end devices often end up in the spotlight for performance improvements$^1$. Third, mobile apps have become increasingly multi-functional, challenging how developers prioritize different parts of a single app.

One of the current industry-standard methods to construct performance metrics in online controlled experiments is to compute a quantile, such as the median or the 90th percentile, over all performance events in a single experiment bucket \cite{liu2019}. At Snap, for example, the in-house experimentation platform runs hundreds of concurrent A/B experiments at any given time, and each experiment in turn reports hundreds of performance metrics calculated in this industry-standard way. In our work improving app performance and investigating the causal relationship between performance and user engagement metrics, we have encountered some puzzling phenomena. In many cases, we observe apparent contradictions between these performance metrics and what user-level engagement metrics show. This motivates us to do investigations and reveal the first pitfall: that industry-standard performance metrics overweight heavily engaged users who are usually also on higher-end mobile devices. In many other cases, we observe performance metrics show significant results in an A/B experiment even though, by design, there should exist no treatment effects on them. We find these false positives arise from a second pitfall: 
treatment effects on user engagement changes sometimes can lead to a gap in distributions of performance event counts per user between treatment and control. This gap can 
cause self-selection bias in performance metrics, which can lead to misleading results. Self-selection bias arises from user behavior changes due to treatment effects, and cannot be avoided by careful experiment design or perfect randomization. Unlike most user engagement metrics which can be imputed by zero when users do not have any events, we cannot do the same for performance metrics \cite{gupta2019}.


    
    

The goal of this paper is to raise industry awareness of these pitfalls in calculating performance metrics and provide practical guidelines for remedying the problems they cause. Here is a summary of our contributions in this paper:
\begin{itemize}
    \item To the best of our knowledge, this is the first work to (1) quantitatively present and explain the difference 
    between industry-standard event-level performance metrics and user-level performance metrics, and (2) demonstrate the importance of using user-level performance metrics to measure mobile apps and how they remedy the issue of over-representing heavy users by event-level.
    
    \item We propose a self-selection Sample Ratio Mismatch (SRM) check for testing whether a performance metric has self-selection bias. We establish the self-selection bias problem as a missing data problem, and apply two methods based on imputation and matching to reduce bias and correct for misleading conclusions.
    
    \item We share discussions on several critical lessons we learned, regarding (1) the scalability and performance of other self-selection bias reduction methods, and (2) the difficulty in computing user-level high percentile performance metrics. These insights should help practitioners avoid pitfalls and problems alike.
\end{itemize}

\section{Preliminaries} \label{sec:prelim}
\subsection{Performance Metrics and Online Controlled Experiments} \label{subsec:perf_ab}

Long app loading and execution times on mobile devices hurt user experiences, drive away user traffic, lower revenue, and cause energy drain \cite{bui2015}. It is thus imperative to construct proper metrics for mobile performance to safeguard against these potential issues. For instance, when a user opens a mobile camera app (e.g. Snapchat), we automatically track how quickly the app opens and how fast one can take photos within the app. These latency values are logged as performance events and uploaded to a central data collection system, resulting in a multitude of individual performance events collected across millions of different mobile devices. 


Online controlled experiments (\cite{kolenikov2016}) have been used as the mantra for data-driven decision making on feature changes and product launches in many mobile and Internet companies. 
In an A/B test, we randomly split users into a treatment group and a control group and observe how metrics of interest move. The Rubin Causal Model \cite{holland86} is commonly used in A/B testing as a statistical framework for causal inference. Let $Y_i(T_i)$ be the potential outcome for $i$-th user, where $T_i = 1$ if the $i$-th user is in the treatment group and $T_i = 0$ if the $i$-th user is in the control group. The Average Treatment Effect (ATE) is defined to be the average of $Y_i(1)-Y_i(0)$, a quantity which is not observable due to the "fundamental problem of causal inference"\cite{holland86}: we cannot observe $Y_i(1)$ and $Y_i(0)$ at the same time. However, an unbiased estimate of ATE can be obtained in the A/B test by calculating the mean difference $\overline{Y_i(1)}-\overline{Y_i(0)}$.

Performance metrics play an important role in online controlled experiments for evaluating and improving new product ideas across mobile and internet companies. Unlike user engagement which is typically measured on the user-level and summarized as a count or a sum in metrics, mobile or website performance is usually measured on the event-level. That is, the current industry standard is to construct \emph{event-level performance metrics} by tracking a representative percentile such as the 50th percentile (the median) or 90th percentile over all performance events. LinkedIn, for instance, monitors the 90th percentile of page loading times for their website \cite{liu2019}. To complement event-level performance metrics, some online services companies like Microsoft also adopt \emph{user-level performance metrics} \cite{li19}: compute a quantile value for each user over all their events. For example, to obtain a user-level 50th percentile performance metric, we first aggregate performance events of each individual user and calculate the median over each user, and then average these values over all users in treatment and control to calculate ATE.


\subsection{Sample Ratio Mismatch (SRM) and Selection Bias} \label{subsec:srm}

Sample Ratio Mismatch (SRM) is one of the major pitfalls of interpreting metrics in online controlled experiments \cite{dirtydozen2017}. Sample Ratio Mismatch (SRM), sometimes also referred as Sample Size Proportion Mismatch (SSPM), is a data quality check that indicates a significant difference between the expected proportions of users between the treatment group and the control group before the experiment and the actual proportions of users observed in the two groups during the experiment.  
SRMs are common in large scale experimentation \cite{chen2018a,zhao2016} and are symptomatic of a variety of data quality issues. For example, SRM issues can be caused by incorrect treatment assignment and lost telemetry \cite{zhao2016}, and can happen in the data processing phase \cite{aleksander2018}. Researchers have provided a taxonomy and rules of thumb for practitioners to diagnose, root cause, and prevent SRMs in online controlled experiments \cite{SRM2019}. SRM check is a practical way to detect a selection bias that arises when a rule other than simple random sampling is used to sample the underlying population \cite{heckman2010}, which invalidates any inference drawn from the experiment \cite{angrist_poschke}



\subsection{Self-Selection Bias} \label{subsec:ss_bias}
Among all types of selection bias, "self-selection bias" is the one in performance metrics that arises from user behavior changes due to treatment effects \cite{dirtydozen2017}. For example, during an A/B experiment, if a user does not use the camera to create images with Snapchat, then for this user, we observe no performance metric value for camera image creation delay, and thus this user is missing from this performance metric calculation. If between control and treatment groups, there are statistically different proportions of users not having that performance metric, then we have self-selection bias for that performance metric. 
Self-selection bias in performance metrics is determined by users, and thus cannot be avoided by having careful experiment design, perfect randomization, or proper counterfactual logging in the data. Unlike most user engagement metrics which can be imputed by zero when users do not have any events, we cannot do the same for performance metrics \cite{gupta2019}.

\section{User-level vs. Event-level} \label{sec:user_vs_event}



Unlike traditional online services companies which use performance metrics mainly as guardrail metrics \cite{evaluation_tutorial2019}, as a mobile app company Snap strives to improve app performance to enhance user experience and engagement.$^1$  Therefore, understanding the correlation between performance metrics and user engagement metrics is crucial for daily decision making at Snap. However, we have observed puzzling results from some A/B experiments: treatment effects on engagement metrics are negative or insignificant while event-level performance metrics improve. In such contradictory cases, experiment designers always struggle to decide whether to deploy feature changes or not. We later discover that this kind of puzzling result only occurs in event-level performance metrics but not in user-level performance metrics. We empirically compare event-level and user-level performance metrics on 152 A/B tests at Snapchat. Table~\ref{Table_1} shows that approximately $10\%$ of the A/B testing results are inconsistent in event-level performance metrics and user-level performance metrics in terms of significance or treatment effect direction. 
\begin{table}[ht]
\captionsetup{font=footnotesize}
\scalebox{0.9}{
\begin{tabular}{|c|c|c|c|}
\hline
\textbf{Event vs. User}  & \textbf{Significant ($+$)} & \textbf{Significant ($-$)} & \textbf{Insignificant} \\ \hline
\textbf{Significant ($+$)} & \textbf{39 (0.47\%)} & 1 (0.01\%) & 178 (2.13\%) \\ \cline{1-4}
\textbf{Significant ($-$)} & 12 (0.14\%) & \textbf{41 (0.49\%)} & 214 (2.57\%) \\ \cline{1-4}
\textbf{Insignificant} & 213 (2.55\%) & 221 (2.65\%) & \textbf{7419 (88.98\%)} \\ \hline
\end{tabular}
}
\caption{Result comparison of the event-level (rows) vs. the user-level (columns), for $8,338$ performance metrics in $152$ A/B tests. Significant ($+$) means $p$-value $<0.05$ with positive estimated ATE. Significant ($-$) means $p$-value $<0.05$ with negative estimated ATE. About $10\%$ of performance metrics are inconsistent between their event-level and user-level.}
\label{Table_1}
\end{table}

\vspace{-20pt}

\begin{figure}[ht]
\begin{center}
\includegraphics[width=0.4\textwidth]{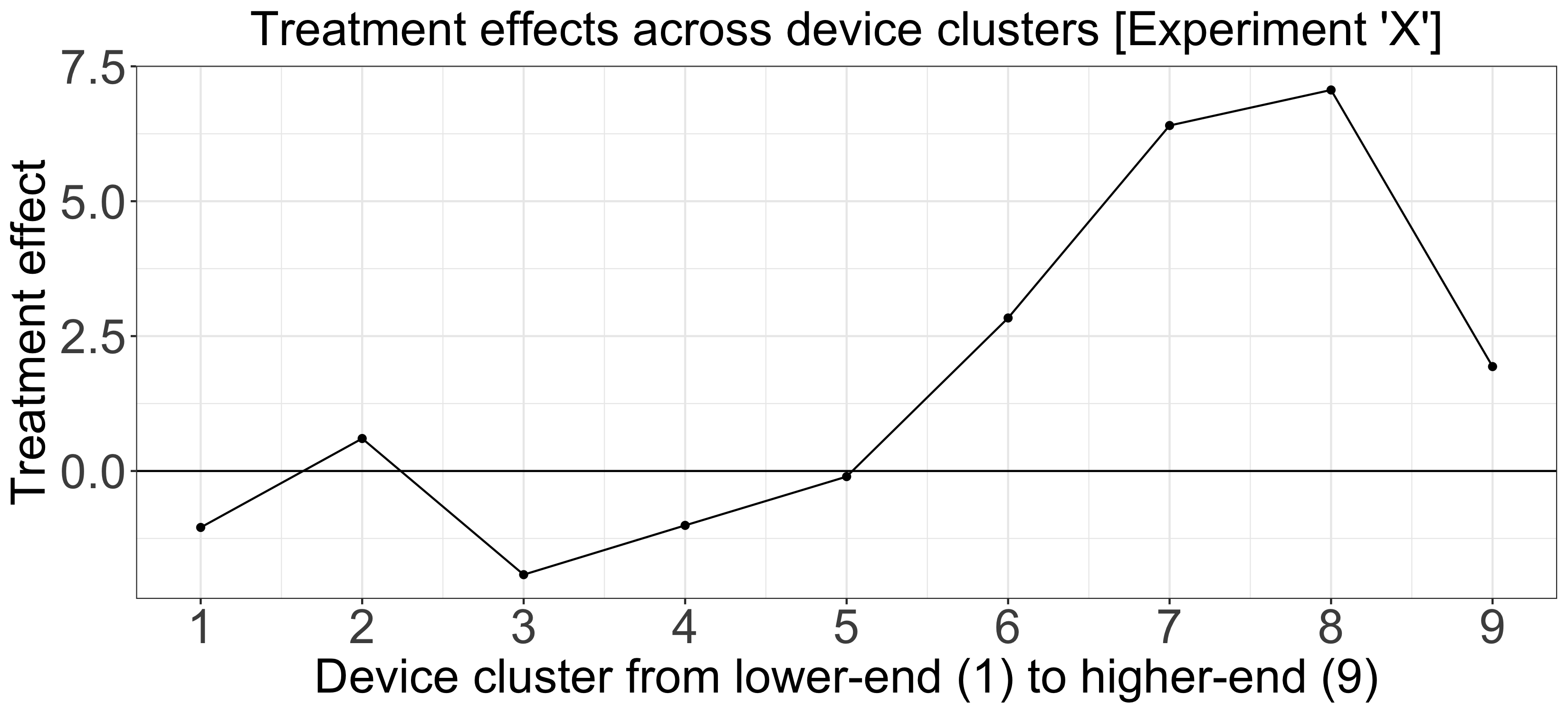}
\caption{ATE across different user groups segmented by device clusters. Users with higher-end devices (right side) benefit from the treatment, while users with mid-end to lower-end devices (left side) do not. In this experiment, the event-level performance metric has a positive ATE $=5.48$ with a significant $p$-value $=0.007$ while the user-level performance metric has an insignificant $p$-value $=0.432$.}
\label{Figure_1}
\end{center}
\end{figure}

\begin{figure}[ht]
\centering
\includegraphics[width=0.4\textwidth,trim={0 0 0 0},clip]{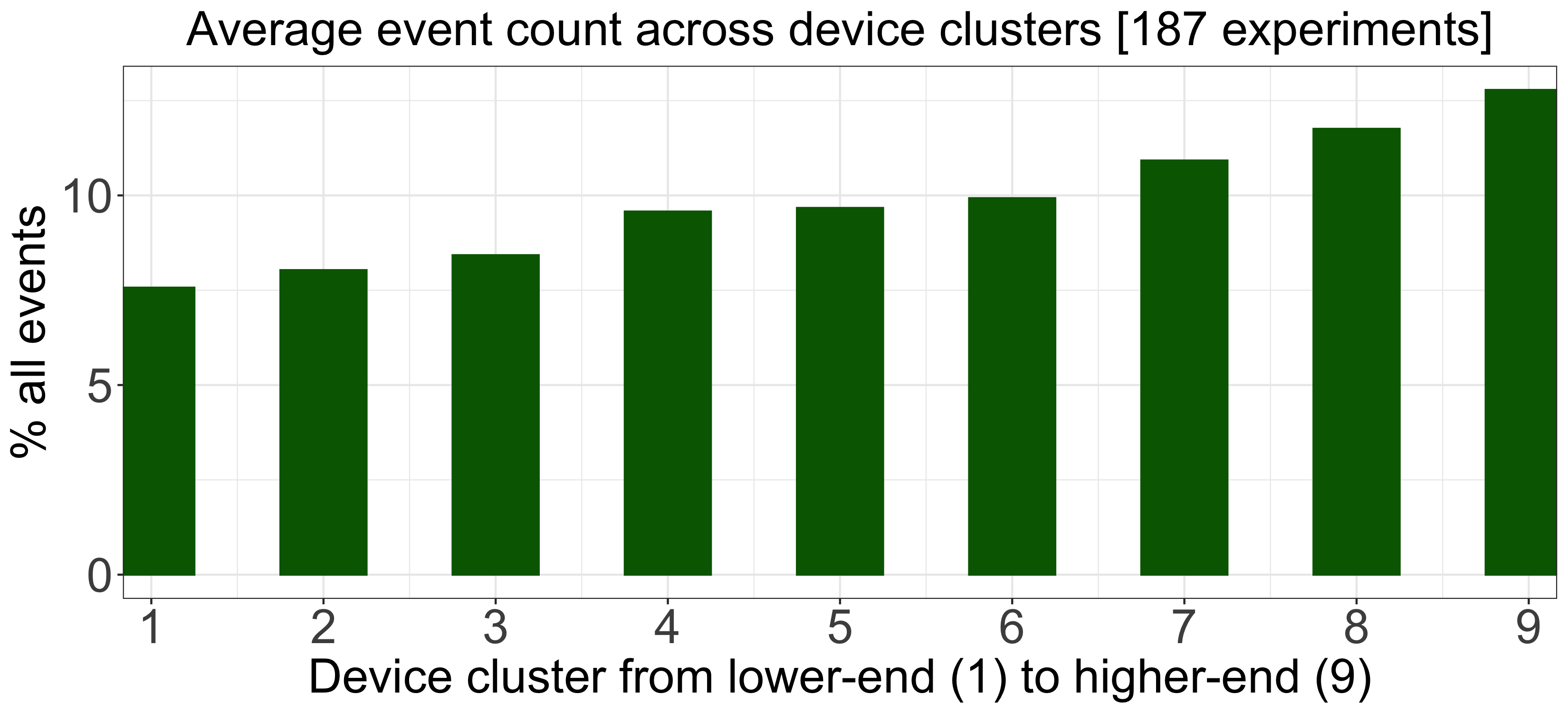}
\caption{
Histogram of relative average event counts per user across different user groups, averaged over 187 experiments. Users with higher-end mobile devices systematically have more events.}
\label{Figure_2}
\end{figure}

\begin{figure}[ht]
\centering
\includegraphics[width=0.4\textwidth,trim={0 0 0 0},clip]{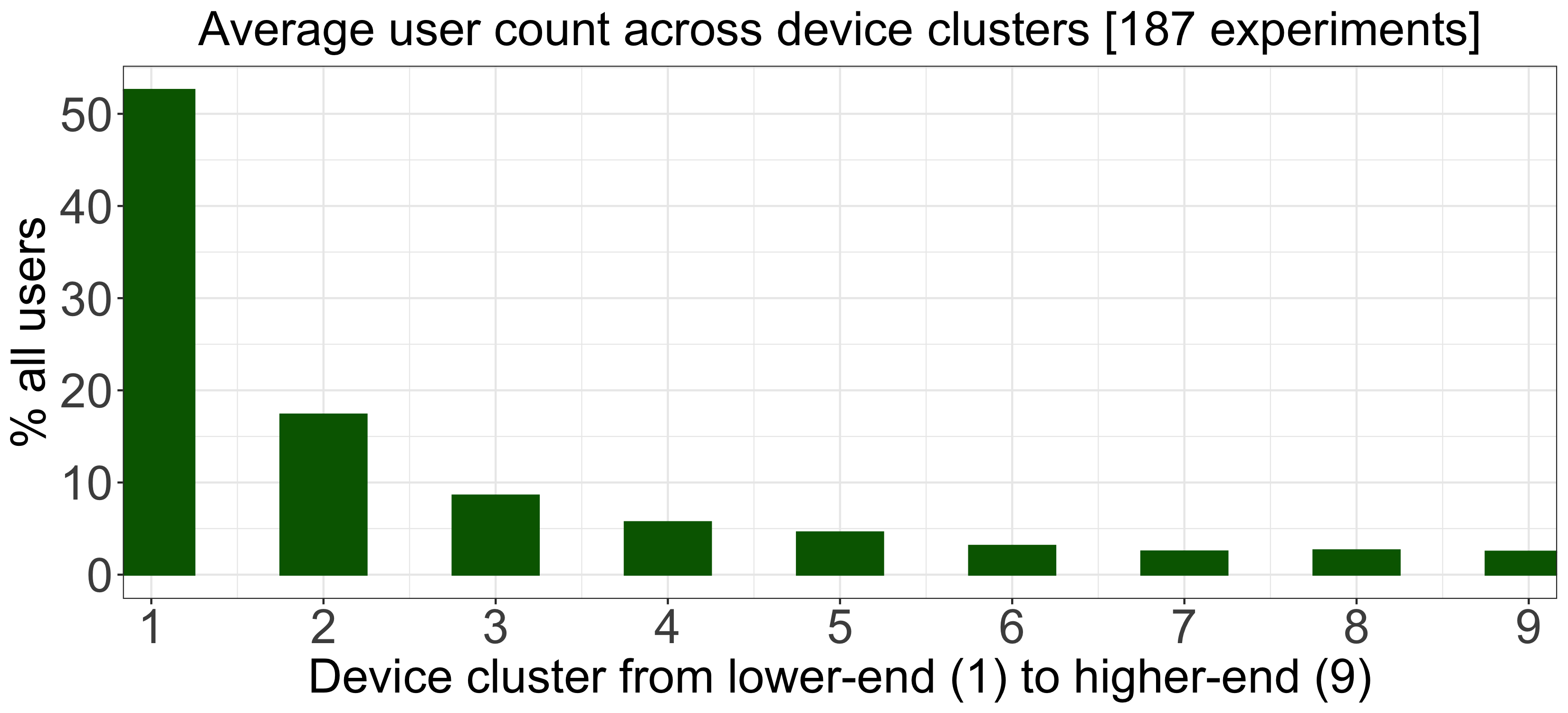}
\caption{
Histogram of relative average user count across different device clusters. There are much more users with lower-end mobile devices (left side) than users with higher-end mobile devices (right side) on market.}
\label{Figure_3}
\end{figure}

After investigation, we find the root cause of the puzzling relationship between user-level metrics and event-level performance metrics is that event-level performance metrics tend to bias towards heavily engaged users while user-level performance metrics weight all users in an experiment equally. Mobile devices are highly heterogeneous, and thus heterogeneous treatment effects are common to be seen in A/B experiments \cite{xie2018}. For example, as Figure~\ref{Figure_1} shows, the estimated average treatment effects vary across different user groups in an experiment "X" and only the users with higher-end mobile devices benefit from the treatment. In addition, users with better mobile devices are likely to engage more with the app and thus have more events as illustrated in Figure~\ref{Figure_2}, indicating that the heavily engaged users almost overlap the higher-end device users. However, since the majority of users use lower-end devices (as shown in Figure~\ref{Figure_3}), if we were to simply focus on event-level performance metrics, we would overlook the majority of those users in our experiments.
Product launch decisions based only on event-level performance metrics are likely to overlook needs of non-heavy users, which is undesirable for expanding into developing or under-developed markets.

Event-level performance metrics overweight heavily engaged users who are usually higher-end device users. Given the strong heterogeneity among mobile devices, treatment effects on higher-end devices can be very different from lower-end ones. However, since majority of mobile users use lower-end devices, event-level performance metrics put these users into our blind spot during experiments. Therefore, to complement event-level performance metrics and to correlate performance improvement with user engagement, we recommend computing both event-level performance metrics and user-level performance metrics to obtain a comprehensive understanding of app performance change in A/B experiments. Event-level performance metrics make sure products maintain good performance on the majority of events, while user-level performance metrics equally weight all users, thus safeguarding the majority of user experiences and better correlating with engagement metrics.

\vspace{-12pt}

\section{Self-Selection Bias Reduction} \label{sec:methods}
In an A/B experiment whose treatment is about app notification, we see another puzzling case: the treatment has nothing to do with app performance according to the experiment design but many performance metrics move significantly on both user-level and event-level. After investigation, we discover that the proportion of users who open the app (i.e. who have values for performance metrics such as app open delay) significantly differs between control and treatment, which leads to self-selection bias in performance metrics, causing false positives as we have mentioned in Section~\ref{subsec:ss_bias}.

Similar to detecting selection bias by using the SRM check \cite{dirtydozen2017}, we propose to detect the self-selection bias of a performance metric by using the self-selection SRM check, which is a proportion test to compare the proportion of users who have the performance event(s) in the treatment group and the corresponding proportion of users in the control group.
Let $p_{T}=Pr\left(M|X_1=\text{treatment}\right)$ and $p_{C}=Pr\left(M|X_1=\text{control}\right)$ where $M$ is an indicator variable for whether the user has any performance event, and $X_1$ is the treatment bucket.
We use a two-sample proportion test in this self-selection SRM check to test the null hypothesis $H_0: p_{T}=p_{C}$ against the alternative hypothesis $H_1: p_{T}\neq p_{C}$. A rejection on the null hypothesis indicates self-selection bias in the given performance metric.
We conduct this self-selection SRM check for 11 performance metrics across 61 A/B experiments showing significant treatment effects on users' active days of using Snapchat, and in more than $20\%$ of these A/B experiments, most performance metrics fail the self-selection SRM check.

Since self-selection bias issue is essentially a missing data problem, in this section, we first introduce different missing data patterns in the literature and discuss the missingness assumption in the scenario of self-selection in performance metrics. We then present two methods, based on imputation and matching, to reduce self-selection bias. These two methods are implemented on user-level performance metrics. In addition, we also list some other methods we have considered for addressing self-selection bias in performance metrics.


\subsection{Assumptions on Missingness} \label{subsec:assump_miss}
The self-selection bias problem in performance metrics is equivalent to a missing data problem, as we only observe performance metric values of users with certain events who are usually a subset of all users exposed in an experiment.


Missing data is a recurring topic of study in data science and it can significantly impact conclusions drawn from data. There are three classical types of missing data patterns. Missing completely at random (MCAR) assumes the values in a data set are missing entirely at random, ensuring subsequent data analysis is valid and unbiased \cite{graham2012}. However, MCAR is usually too strong an assumption to be true in practice. Missing at random (MAR) assumes missingness can be fully accounted for by variables with complete information \cite{raghunathan2016}. It is a weaker assumption than MCAR, though uncheckable, and thus requires reasonable justification. Under the assumption of MAR, many methods are able to produce asymptotically unbiased estimates of parameters estimated from data (e.g. mean of the data). Another type of missingness is missing not at random (MNAR), where the data is neither MAR nor MCAR. MNAR requires more conditions than MAR does for consistent parameter estimation. 

Based on our experience with missing data in performance metrics, we have three observations about the likelihood of a user not having any event for a given performance metric:
\begin{enumerate}[label=(\Roman*)]
    \item A treatment in an online controlled experiment may affect how many users use the app. For example, the treatment causes a system bug which prevents Android users from logging into the app, causing more missingness in the treatment group than in the control group. This assumption may or may not hold depending on the experiment design.
    
    \item Missing rates differ statistically among different device models. We have checked in real data to see that users with lower-end devices and worse mobile app experience are more likely to have no performance events in experiments. 
    
    \item Historically less engaged users are more likely to have no performance events in an experiment. We have checked this assumption in real data to ensure that the missing rate of high-engaged user groups is much lower than the missing rate of low-engaged user groups.
\end{enumerate}
These observations ensure the missingness is not MCAR. 
Instead, based on these observations it is reasonable to assume that the missingness is close to MAR and can be mostly accounted for by three key factors: the treatment bucket, the device model, and the pre-treatment engagement level of users, i.e. 
 \vspace{-3pt}
\begin{equation} \label{mar}
    Pr\left(M | Y, X_1, X_2, X_3\right)
   \approx Pr\left(M | X_1, X_2, X_3\right), 
\end{equation}
where $M$ is an indicator variable for whether a user has any event for the performance metric, $Y$ is the performance value, $X_1$ is the treatment bucket, $X_2$ is the device model type, and $X_3$ is the pre-treatment engagement level.
The interpretation of \eqref{mar} is: the probability of missing performance metric data primarily depends on users' treatment buckets, device model types, and pre-treatment engagement levels. Within a cluster of users having the same treatment id, device model and pre-treatment engagement level, the probability of missing data on the performance metric is almost unrelated to the performance values. 

Although MAR cannot be verified statistically against MNAR \cite{little1986}, it is a practical assumption to be made given the observations (I) -- (III). In theory it is possible for many other variables to also partially account for the missingness of performance metrics, but in practice we recommend considering only a small number of main factors out of scalability concerns. We find no significant improvement in results for most experiments at Snap by adding more variables. Practitioners in different companies may vary on their choices of variables. Based on the assumption of missingness, we propose the following two methods to reduce self-selection bias in user-level performance metrics.


\subsection{Method 1: Imputation} \label{subsec:imputation}
The first method we propose is based on imputation. Imputation has always played an important role in the study of missing data and so researchers have developed many methods.
`Mean substitution' is one of the easiest and fastest ways of imputation \cite{little1992}, which simply uses the mean to fill in missing values. Random imputation is another scalable imputation method which replaces missing values of a variable with some randomly sampled values of that same variable from another sample \cite{lanning2003, altmayer2002}. Moreover, a commonly used imputation method is to replace missing values by the predicted values from a regression model using other predictor variables \cite{schafer1997, faraway1997}. In addition, many researchers also use EM algorithms for imputation \cite{little1986}. Simulation studies in \cite{enders2002} suggest that EM algorithms may perform better than the mean substitution and the random imputation methods. However, these iterative algorithms are computationally expensive. As deep learning and neural networks have achieved star status in the popular press, imputation methods using deep learning and neural networks have also been developed recently \cite{maiti2008, smieja2018, biessmann2019, arisdakessian2019}. A common major drawback of these methods is that they can be very slow when datasets are large. Besides the aforementioned single imputation methods, multiple imputation \cite{rubin1987} is also popular. However, multiple imputation does not scale to large datasets as well as single imputation does.

At Snap, we have thousands of A/B experiments running every day and most experiments involve tens of millions of users. For the scale of this data, a good imputation method requires not only high accuracy but also high scalability.
Considering this along with the missingness assumption on our data \eqref{mar}, we propose to implement the following user-level imputation method:
\begin{algorithm}[ht] 
\caption{Imputation}
\label{alg:loop}
\begin{algorithmic}[1]
\Require{The original user-level performance metric ($Y_{\text{origin}}$); Treatment indicator variable ($X_1$); Device model type variable ($X_2$); Pre-treatment engagement level variable ($X_3$).} 
\Ensure{The user-level performance metric with imputed values ($Y_{\text{impute}}$)}
\Statex
\Function{Imputation}{Input}
  \State {Split users into $G$ subgroups based on $X_1$, $X_2$, and $X_3$.} 
\For{$g \gets 1$ to $G$}                  
    \State{In the $g$-th subgroup, use the $Y_{\text{origin}}$ values of the non-missing users to construct an empirical distribution $F_g$.}
    \State{Randomly sample values with replacement from $F_g$ to fill in the missing values of $Y_{\text{origin}}$ in the $g$-th subgroup.}
  \EndFor
  \State \Return {The user-level performance metric after imputation}
\EndFunction
\end{algorithmic}
\end{algorithm}

%
%
%

In Algorithm 1, the number of subgroups depends on the number of device models and the number of pre-treatment engagement levels. On one hand, given the heterogeneity of device models as shown in Section~\ref{sec:user_vs_event}, having more specific subgroups may increase the accuracy of imputed values. On the other hand, splitting users into too many subgroups decreases the scalability of our method and may result in many groups with very few non-missing users, thus decreasing the accuracy of imputation. Hence, we keep only device models with at least 10,000 users and classify all other models as "Other." As for the pre-treatment engagement level, we first count each user's number of performance events in one week before the experiment start, and then divide the users into a high-engagement group (count $>$ mean of counts) and low-engagement group (count $\leq$ mean of counts).

In addition to its easy and straightforward implementation, we choose to use random imputation in our algorithm because, unlike mean substitution, it yields variance in imputed values which is useful for hypothesis testing. 
It is also much faster than other imputation methods for large datasets. 

If the equality in \eqref{mar} holds exactly, then imputed values will be generated from the correct distributions and asymptotic unbiasedness in parameters (e.g. mean, median, etc.) is achieved. Moreover, we impute missing values in control group and treatment group separately, which preserves any treatment effects on performance metrics. In reality, we do not expect the equality in \eqref{mar} to always hold, so we cannot claim to completely correct self-selection bias. But as long as \eqref{mar} is close to the truth, our imputation method reduces self-selection bias. We evaluate the performance of our methods via simulations and empirical studies in Sections~\ref{subsec:simulation} and ~\ref{subsec:real_study}.

\vspace{-4pt}

\subsection{Method 2: Matching} \label{subsec:matching}

The second method is based on matching. The idea of matching is to balance the distributions of observed variables between treatment and control groups. It is widely used for selection bias problems in observational studies across diverse disciplines such as statistics \cite{rosenbaum2002, rubin2006}, economics \cite{imbens2004}, political science \cite{ho_imai_king_stuart_2007}, sociology \cite{morgan_harding_2006}, and epidemiology \cite{Epidemiology2006}.
Among all well-developed matching methods, "Exact Matching" matches treatment units with control units who share exactly the same values in all observed variables. 
Exact matching is ideal in many aspects \cite{ho_imai_king_stuart_2007}, yet
finding an identical twin for each treated unit is difficult in high-dimensional or continuous-valued data. 
To deal with this, "Coarsened Exact Matching" \cite{iacus_king_porro_2012} allows for close matching of continuous variables, while approaches based on propensity scores \cite{rosenbaum_rubin_1983} work for high-dimensional data \cite{imai_ratkovic_2014}. "Entropy Balancing" is another recent matching method \cite{hainmueller2012} that directly estimates weights to balance observed variables between treatment and control groups. 

In order to ensure good scalability and accuracy, we implement "Exact Matching" based on users' device models and pre-treatment engagement levels as shown in Algorithm 2. After calculating weights for matching, we use weighted t-tests to compare the user-level performance metrics between the treatment group and the control group. 

\begin{algorithm}[ht] 
\caption{Matching}
\label{alg:loop2}
\begin{algorithmic}[1]
\Require{The original user-level performance metric ($Y_{\text{origin}}$); Treatment indicator variable ($X_1$); Device model type variable ($X_2$); Pre-treatment engagement level variable ($X_3$).} 
\Ensure{The weighted user-level performance metric ($Y_{\text{match}}$)}
\Statex
\Function{Matching}{Input}
  \State {
  Extract the non-missing values $Y_{\text{NM}}$ out from $Y_{\text{origin}}$. 
  }
  \State {
  Split users with $Y_{\text{NM}}$ into $G$ group based on $X_2$ and $X_3$. 
  }
\For{$g \gets 1$ to $G$}                  
    \State{In the $g$-th subgroup, denote the number of treated users (i.e. $X_1=1$) to be $N^T_g$ and the number of control users (i.e. $X_1=0$) to be $N^C_g$, and compute the ratio $\frac{N^T_g}{N^C_g}$.}
    \State{Assign weights $\{1, \frac{N^T_g}{N^C_g}\}$ to the $Y_{\text{NM}}$ values of $\{treated, control\}$ users.}
  \EndFor
  \State \Return {The weighted performance metric after matching}
\EndFunction
\end{algorithmic}
\end{algorithm}

In our practice, both device models and pre-treatment engagement levels are categorical. We have also tried matching based on some continuous variables which might affect app performance such as app versions and OS versions, by using propensity score matching \cite{rosenbaum_rubin_1983} to handle these continuous variables. None of them significantly reduces bias in the cases we have examined. Practitioners in other companies may choose different variables or matching methods for their own needs.

\subsection{Imputation versus Matching} \label{subsec:imputation_vs_matching}

The above imputation and matching methods are based on the same assumption, which is described in Equation \eqref{mar}: for a cluster of users with the same experiment bucket, device model, and pre-treatment engagement level, the distribution of a performance metric's missing values should match that of its observed ones. 
If this equation holds exactly, both methods can achieve asymptotically unbiased estimates of the parameters such as the mean and median. In Section~\ref{subsec:simulation}, we show via simulations that both methods are able to obtain close-to-truth estimated ATEs under the assumption of this equation. However, these two methods focus on different populations, which affects how well estimated treatment effects can generalize to different user groups. In an A/B experiment, imputation targets all users exposed in this experiment, while matching targets only users who have the performance metric values observed in the experiment. 

The Equation~\eqref{mar} may not always hold in real datasets, as the MAR assumption is untestable in general unless additional distributional assumptions or instrumental variables are given (\cite{breunig2019, manfred2006}). So both methods do not completely correct self-selection bias. Since we usually do not know the ground truth treatment effects in real A/B experiments, we recommend using both methods to cross-check results after self-selection bias reduction.



\subsection{Simulation Studies} \label{subsec:simulation}
We first conduct simulation studies to evaluate the performance of our proposed methods. To mimic real distributions of the user-level performance metrics at Snap, we randomly select $3$ millions users and use their performance metric data in our simulation studies. We randomly split the users into treatment and control groups so that initially there is no treatment effect. To generate treatment effects for some of our simulation scenarios, we assume that different device models have different treatment effects. Therefore, we first generate a value of $\mu \sim \text{Uniform}(a, b)$ for each device model, where $a$ and $b$ are chosen to control the sizes of treatment effects. Then, we add values sampled from a Gaussian distribution $N(\mu, \sigma)$ to users in the treatment group based on their device models, where we set $\sigma=0.3$ in the following simulation studies. This establishes a ground truth for the average treatment effect (ATE).

The self-selection bias in our simulated datasets is generated according to the Equation \eqref{mar} in Section~\ref{subsec:assump_miss}. The probability of a given user not having a value of the performance metric $Y$ is defined to be
\begin{equation} \label{missingness}
    \vspace{-4pt}
    Pr\left(M | X_1, X_2, X_3\right)= expit\left(\alpha_0+\alpha_1 X_1+\alpha_2 X_2+\alpha_3 X_3\right), 
\end{equation}
where $M$ is an indicator variable for whether a user is missing the performance metric $Y$, $X_1$ is the indicator variable of the treatment bucket, $X_2$ is some dummy variable for the device model type, $X_3$ is the pre-treatment engagement level. Increasing $\alpha_0$ decreases the overall missing rate, increasing $\alpha_1$ amplifies the difference in missing rates between the treatment group and the control group, and the final parameters $\alpha_2, \alpha_3$ are set to adjust the effect of device model type and pre-treatment engagement level on the missingness.

We denote the level of self-selection SRM to be 
\begin{equation*}
  \vspace{-4pt}
    \Delta_{\text{miss}}=\% \text{missing in treatment} - \% \text{missing in control}. 
\end{equation*}
We consider three simulation setups where all results are averaged over 100 iterations of the data. In the first setup, we fix the true ATE introduced into the treatment group to be $3$, set $\alpha_0, \alpha_2, \alpha_3$ constant, and only adjust $\alpha_1$ in \eqref{missingness} to vary $\Delta_{\text{miss}}$. 
As the Figure~4($A_1$) shows, as $\Delta_{\text{miss}}$ increases, the ATE estimated directly from the original data grows further away from the true ATE, while both imputation and matching obtain estimated ATE values close to the truth at all $\Delta_{\text{miss}}$'s. Since the true ATE is positive, if hypothesis testing gives either an insignificant $p$-value (i.e. $p$-value$\ge 0.05$) or a significant $p$-value with a negative estimated ATE, we would consider this a false conclusion. We define the false conclusion rate (FCR) to be 
\begin{equation*}
\vspace{-3pt}
FCR=\frac{\text{Number of false conclusions made}}{\text{Total number of hypotheses tested}}.    
\end{equation*}
Figure~4($A_2$) shows that both imputation and matching make no false conclusions in this simulation setup. Without a correction method, we see the $FCR$ equal $1$ when $\Delta_{\text{miss}}$ reaches a certain level.

In the second simulation setup, we introduce no treatment effect to the treatment group, setting the true ATE equal to $0$. In this case, any significant $p$-value indicates a false positive and we compute the false positive rate (FPR) for each method. As Figure~4($B_1$) and Figure~4($B_2$) show, both imputation and matching consistently estimate the ATE around zero and almost never yield false positives. In contrast, hypothesis testing without a correction method performs very poorly.

In the third simulation setup, we fix $\alpha_0, \alpha_1, \alpha_2, \alpha_3$ to make $\Delta_{\text{miss}}=2\%$, and vary the size of the true ATE. We see from the Figure~4($C_1$) that our proposed methods again perform very well in estimating the ATE, with imputation slightly outperforming matching in reducing bias. Given that all estimated ATE values here are positive, any insignificant $p$-values indicate a false negative. The Figure~4($C_2$) shows that our proposed methods have low false negative rates (FNR) even when the true ATE is very small. It is not surprising to see that the FNR from using the missing data without a correction method decreases as the true ATE grows, because if the true ATE is significantly large, a $2\%$ self-selection SRM level may not be powerful enough for self-selection bias to produce wrong conclusions, especially with a sample size in only the low millions.

\begin{figure*} \label{fig:sim}
\begin{center}
    \includegraphics[width=0.33\textwidth]{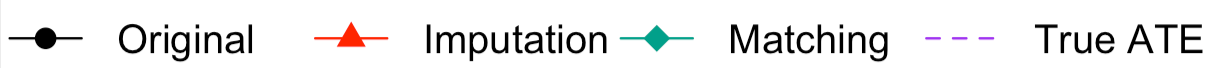}
\end{center}
    \includegraphics[width=0.33\textwidth]{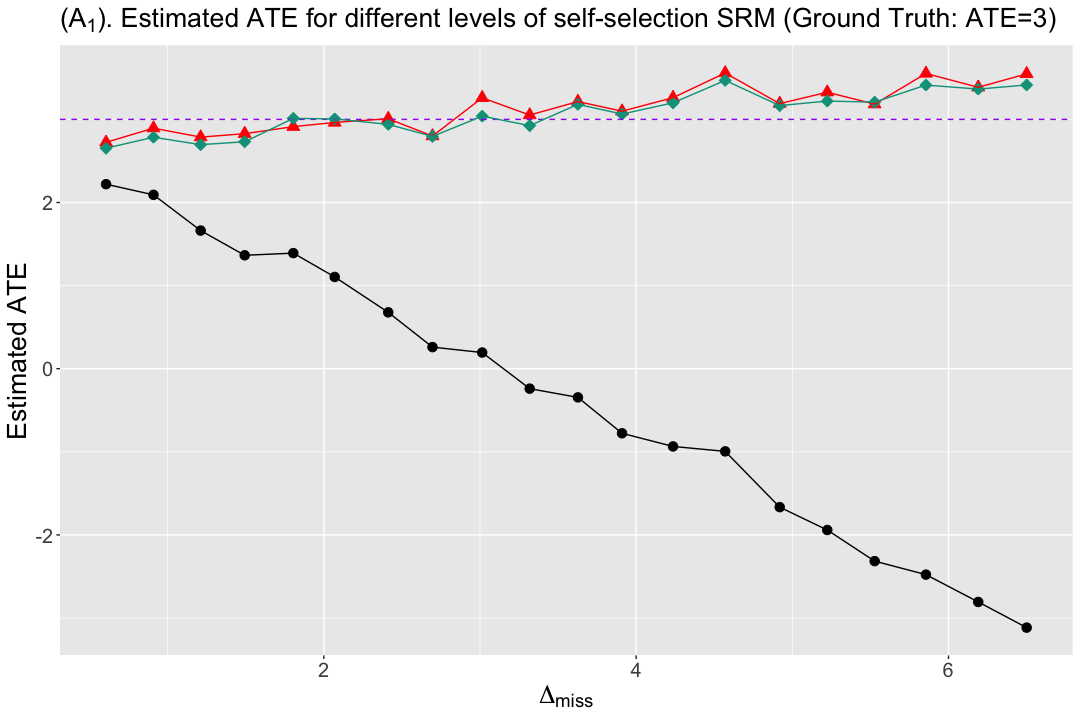} \includegraphics[width=0.33\textwidth]{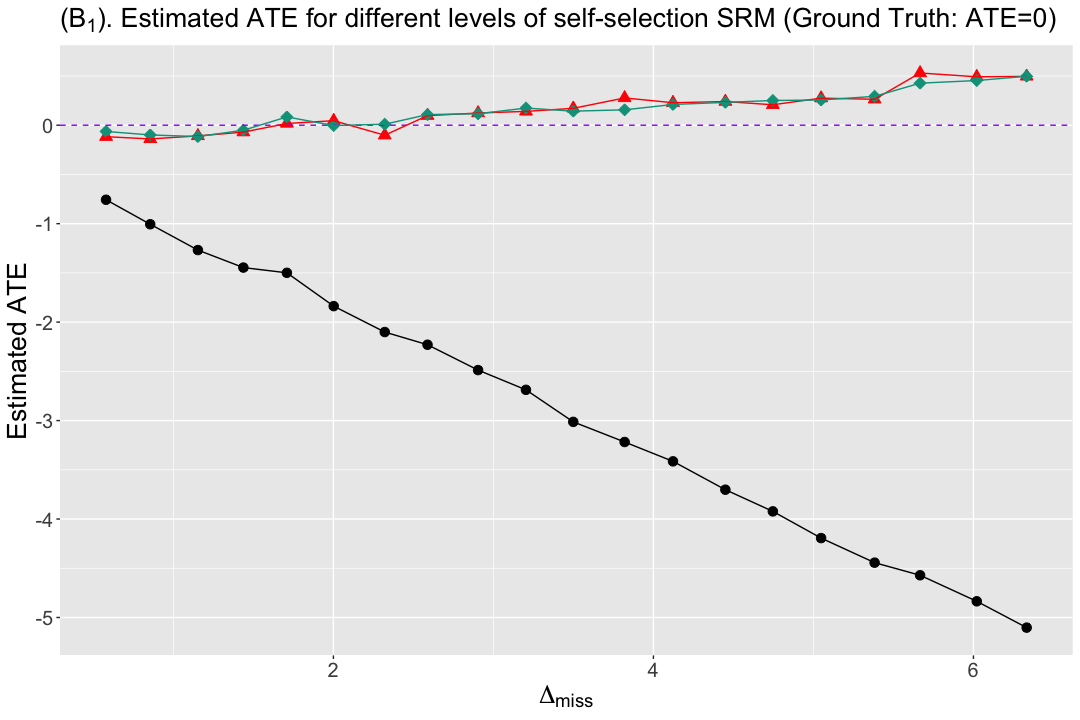}\hfill
	\includegraphics[width=0.33\textwidth]{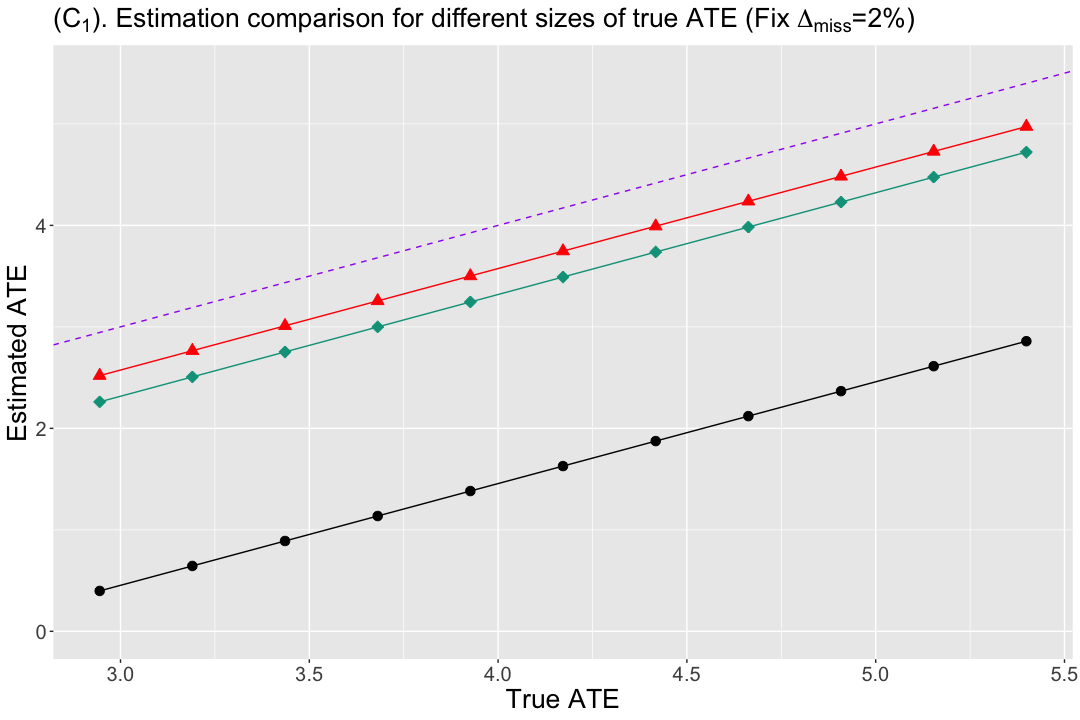} \includegraphics[width=0.33\textwidth]{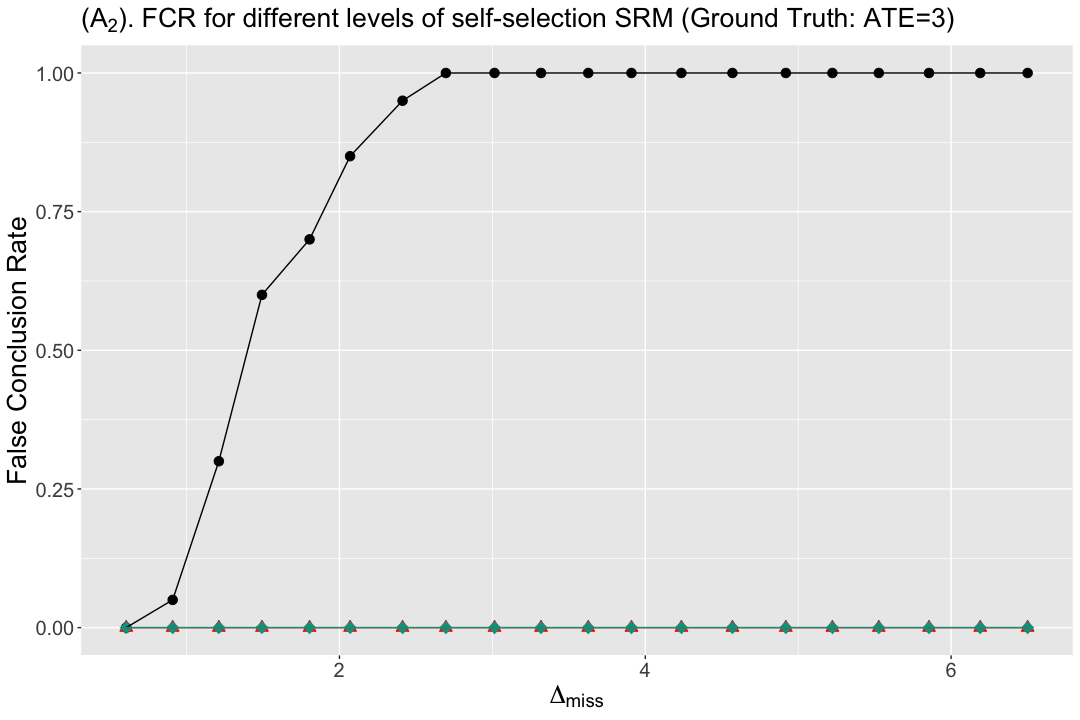}\hfill
	\includegraphics[width=0.33\textwidth]{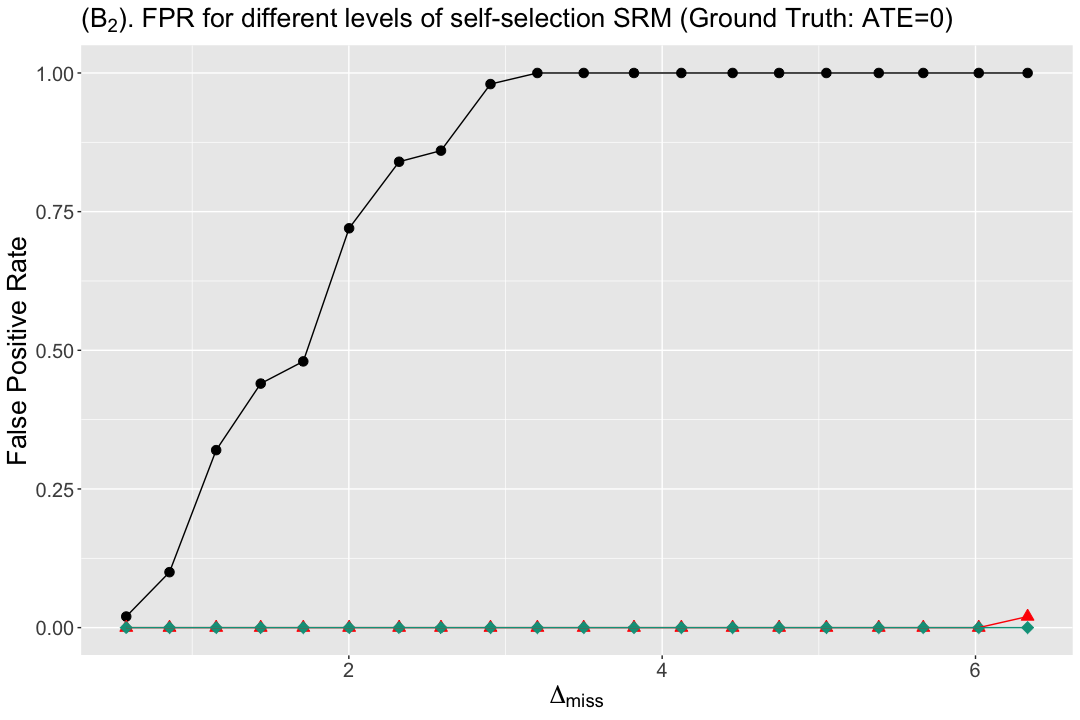} \includegraphics[width=0.33\textwidth]{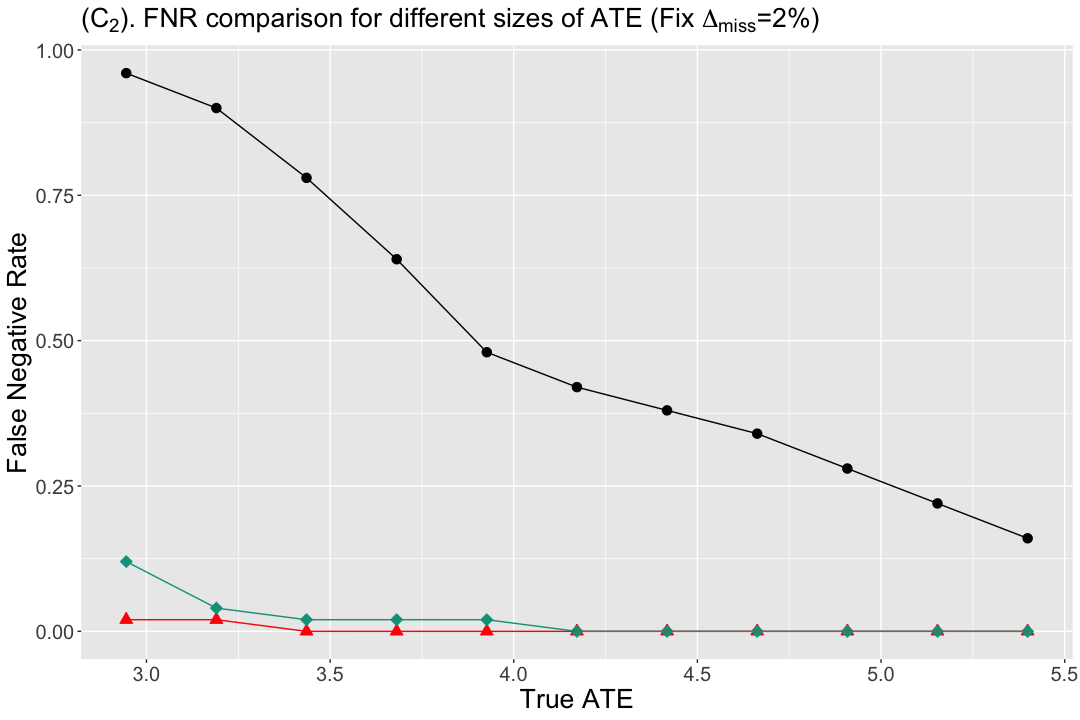}\hfill
	\caption{Simulation results for imputation, matching, and without correction on user-level performance metrics. For ($A_1$) and ($A_2$), we fix the true ATE to be $3$ and vary $\Delta_{\text{miss}}$. For ($B_1$) and ($B_2$), we fix the true ATE to be $0$ and vary $\Delta_{\text{miss}}$. For ($C_1$) and ($C_2$), we fix $\Delta_{\text{miss}}=2\%$, and vary the size of true ATE.  }
\end{figure*}


\vspace{-4pt}
\subsection{Empirical Results} \label{subsec:real_study}

In addition to simulation studies, we also evaluate the performance of our proposed methods extensively on real A/B experiments at Snap. The difficulty of evaluating methods on a real A/B experiment is that we usually can neither confirm the existence of treatment effects nor ascertain the ground truth treatment effects to performance metrics in an experimental context. Fortunately, we can exploit experiments that by design, should show no treatment effects to performance metrics.
These experiments essentially serve as A/A tests, as it is reasonable to assume that true treatment effects are close to 0. If performance metrics in these experiments exhibit self-selection bias and indicate statistically significant mean differences between treatment and control groups, then we can test whether our methods can successfully correct these false positives.

We look at the A/B experiment~1 where the volume of notifications changes in the treatment group. According to those who designed the experiment, there should be no treatment effect to the performance metric $X$. However, we observe a significant $p$-value for this metric. In this A/B experiment, $67.79\%$ of users in the treatment group have values of $X$ while the corresponding proportion in the control group is $68.24\%$. This fails our self-selection SRM check with a $p$-value$<0.001$. We compute the user-level performance metric $X$ and the test result turns out to be significantly negative, as shown in A/B experiment 1 of Table~\ref{tab:experiment} . This is an example of a false positive in user-level performance metrics. After applying imputation and matching methods described earlier, we obtain insignificant p-values from both methods, thereby correcting this false positive.

In the A/B experiment~2 where treatment uses a different machine learning algorithm for notifications, and where we expect no impact on performance metric $Y$. 
Here, $54.25\%$ of users in the treatment group have values of $Y$ while the proportion of users having $Y$ in the control group is $54.35\%$. This fails our self-selection SRM check with a $p$-value$<0.001$. We see from the A/B experiment 2 of Table~\ref{tab:experiment} that our proposed methods again correct the false positive in this A/B experiment.

These two examples have demonstrated that our proposed methods are able to correct false positives in real A/B experiments. In the A/B experiment~3 the treatment should not affect the performance metric $Z$ by design, while $Z$ fails our self-selection SRM check with a $p$-value$<0.001$ ($85.24\%$ of users non-missing in treatment vs. $85.05\%$ in control). Although there is still self-selection bias, the user-level $p$-value for this performance metric $Z$ is insignificant and so does not produce misleading conclusions. As shown in A/B experiment 3 of Table~\ref{tab:experiment}, our methods also yield insignificant $p$-values, indicating neither method over-corrects for self-selection bias.

As mentioned in Section~\ref{subsec:assump_miss}, our proposed methods rely on the missingness assumption \eqref{mar}. Since the Equation \eqref{mar} does not always hold in practice, our methods do not completely correct self-selection bias. Therefore, in some A/B experiments our methods may not be able to correct false positives. In the A/B experiment~4, treatment touches chat notifications and the experiment owner confirms no treatment effect on the performance metric $W$. This performance metric fails our self-selection SRM check with a $p$-value$<0.001$ ($89.32\%$ of users having performance events in treatment vs. $89.39\%$ in control). The user-level $p$-value without a correction method is significant, and both our methods fail to correct this false positive. However, as shown in A/B experiment 4 of Table~\ref{tab:experiment}, both methods still reduce bias term by over $20\%$, assuming a true treatment effect of 0. 

\begin{table}[ht]
    \centering
    \scalebox{0.8}{
    \begin{tabular}{ |c|c|c|c| } 
    \hline
    \multicolumn{4}{|c|}{A/B experiment 1 ($\Delta_{\text{miss}}=0.45\%$, self-selection SRM $p$-value$<0.001$)} \\
    \hline
            & Without Correction & Imputation & Matching \\ 
    \hline
    mean difference & -2.11 & 0.7 & -0.11  \\ 
    \hline
    p-value & \color{red}{0.03}\color{black} & 0.428 & 0.912  \\ 
    \hline
     \multicolumn{4}{|c|}{A/B experiment 2 ($\Delta_{\text{miss}}=-0.1\%$, self-selection SRM $p$-value$<0.001$)} \\
     \hline
            & Without Correction & Imputation & Matching \\ 
    \hline
    mean difference & 1.87 & 0.94 & 0.26  \\ 
    \hline
    p-value & \color{red}{0.022}\color{black} & 0.167 & 0.748  \\ 
    \hline

    \multicolumn{4}{|c|}{A/B experiment 3 ($\Delta_{\text{miss}}=-0.19\%$, self-selection SRM $p$-value$<0.001$)} \\
    \hline
            & Without Correction & Imputation & Matching \\ 
    \hline
    mean difference & -0.77 & -0.3 & -0.31  \\ 
    \hline
    p-value & 0.594 & 0.825 & 0.828  \\ 
    \hline
    \multicolumn{4}{|c|}{A/B experiment 4 ($\Delta_{\text{miss}}=-0.07\%$, self-selection SRM $p$-value$<0.001$)} \\
    \hline
            & Without Correction & Imputation & Matching \\ 
    \hline
    mean difference & 2.32 & 1.89 & 1.53  \\ 
    \hline
    p-value & \color{red}<0.001 & \color{red}<0.001 & \color{red}0.001 \\ 
    \hline
    \end{tabular}
    }
    \caption{Mean differences and $p$-values from using performance metrics in several A/B experiments without correcting self-selection bias (left), after imputation (middle), and after matching (right). }
    \label{tab:experiment}
    \vspace{-20pt}
\end{table}

\vspace{-4pt}
\section{Discussion} \label{sec:discussion}
\subsection{Guidelines for Practitioners} \label{subsec:recom_prac}
Based on our experience at Snap, we recommend the following guidelines to mobile app developers on measuring performance metrics in their online controlled experiments:

\begin{itemize}
    \item \textbf{Step 1}: For a given performance metric, compute both user-level and event-level.
    
    \item \textbf{Step 2}: Detect whether there is self-selection bias in the performance metric by using self-selection SRM check. Alert if $p-$value $<0.001$.
    
    \item \textbf{Step 3}: If there is a self-selection SRM alert in this performance metric, then apply both imputation and matching methods to reduce the self-selection bias.
    \vspace{-5pt}
\end{itemize}

\subsection{Scalability} \label{subsec:scalability}
We have tested the scalability of our methods with different combinations of $N=$ number of users and $G=$ number of subgroups. We implemented our methods in a toolkit at Snap and ran this toolkit on a data set with $N=6$  million and $G=100$ for 100 iterations. For both methods, the tasks have been finished under one hour on an Apple 15" MacBook Pro 2.6GHz Intel Core i7. We believe that more computation power would speed this up even further. 

\subsection{Other Methods} \label{subsec:other_methods}
 We have also tried several other methods to reduce self-selection bias for performance metrics. When analyzing the full user population in an A/B experiment, post-stratification and calibration weighting \cite{kolenikov2016}, which are widely used to adjust survey attrition, yield levels of performance and scalability similar to those of imputation. When analyzing only users with observed performance metric values in an A/B experiment, regression adjustment \cite{haneuse2009} and double robustness \cite{kang_schafer2007} are two alternatives to matching. We have also tried the Heckman correction model \cite{heckman1979} without the exclusion restriction to reduce self-selection bias. However, it performs poorly since it depends on a very restrictive assumption: errors of the selection model and the outcome variable are jointly Gaussian distributed. This assumption rarely holds empirically. Although there are methods to relax this assumption (\cite{newey1990}, \cite{wojtys2016}, \cite{zhelonkin2015}), its scalability is poor. 

In addition to the aforementioned point estimation methods, Horowitz-Manski bounds \cite{horowitz_manski2000} and Lee bounds \cite{lee2009} are commonly used to bound a treatment effect with an interval. However, both methods are of limited use for our particular analysis since they either yield bounds too wide to be informative, or carry assumptions too restrictive for industry settings.


Self-selection bias affects performance metrics on both the user-level and the event-level. 
We have considered methods to reduce self-selection bias in event-level performance metrics, but their scalability becomes challenging. We leave this to future research.

\subsection{User-Level High Percentile Performance Metrics} \label{subsec:high_percentile}
It is critical for mobile apps to guard against bad performance, so high percentiles (e.g. P75, P90) are commonly used in performance metrics \cite{evaluation_tutorial2019}. In addition to the major pitfalls we have discussed in Section~\ref{sec:user_vs_event} and Section~\ref{sec:methods}, there is another challenge in computing user-level high percentile performance metrics. For example, when a user only records two events for a P90 performance metric, it is difficult to derive an accurate value of this user's P90 performance metric from only two data points. From our simulations, user-level high percentile performance metrics get underestimated for users with very few events. 

If an experiment does not modify the distribution of performance event counts between treatment and control groups, then this issue would not affect A/B test results and our proposed methods would still work in reducing self-selection bias. However, if the treatment changes how many events users record, then this issue of underestimation in user-level high percentile performance metrics may lead to misleading results. One possible solution is to impute more accurate values of user-level P90 performance metrics for users with very few events, though this introduces additional assumptions.

\vspace{-4pt}
\section{Summary} \label{sec:summary}
In this paper, we discuss a couple of major pitfalls in the current industry-standard method of calculating performance metrics of mobile apps in online controlled experiments: one is caused by the high heterogeneity of both mobile devices on the market and user engagement with mobile apps, and the other is caused by the user self-selection bias happening during the treatment of experiments. In order to evaluate and remedy the biases caused by these pitfalls, we quantitatively compare the user-level vs. event-level of thousands performance metrics, introduce self-selection SRM alert, and propose imputation and matching methods to remedy the self-selection bias. As shown through both simulation and real A/B experiments, our proposed methods provide robust, scalable ways to reduce biases and remedy misleading results. A general guideline for industry practitioners is also provided.

\section{Acknowledgement} \label{sec:acknowledge}
We thank the anonymous reviewers and the editor for helpful comments on this work. We thank the engineering team at Snap for providing valuable insights and data sets of performance metrics in online controlled experiments.

\newpage
\bibliographystyle{plain}
\bibliography{references}
\end{document}